\documentclass{article}

\usepackage{graphicx}
\usepackage[round]{natbib}
\usepackage{epsfig}
\usepackage{graphicx,amsmath,amsfonts,latexsym}
\title{
CMIP3 ensemble spread, model similarity, and climate prediction uncertainty
}
\author{
Stephen Jewson (RMS) \footnote{\emph{Correspondence email}: \texttt{stephen.jewson@rms.com}} \\
and \\
Ed Hawkins (NCAS-Climate, University of Reading)
}

\begin{document}
\maketitle

\begin{abstract}
The CMIP3 multi-model ensemble spread most likely underestimates the real model uncertainty in future climate predictions because of the similarity, and shared defects, of the models in the ensemble.
To generate an appropriate level of uncertainty, the spread needs inflating.
We derive the mathematical connection between an assumed level of correlation between the model output and the necessary inflation of the spread,
and illustrate the connection by making temperature predictions for the UK for the 21st century using four different
correlation scenarios.
\end{abstract}

\section{Introduction}
The IPCC report~\citep{IPCC4} bases many of its conclusions about future climate on a set of integrations of numerical climate models, which are described in~\citet{cmip}.
These climate models all show future temperatures increasing, with some differences between the models as to exactly how much the temperatures will increase.
The mathematical interpretation of this model ensemble is somewhat difficult, for a number of reasons.
Firstly, the individual models are likely overfitted, since they have many parameters, and much effort has been expended in tuning the models to capture past climate.
This means that the goodness of fit of the models to past climate cannot be taken as an indication of
their ability to predict future climate.
Secondly, the models should not be considered to be independent random samples from some space of all possible models. Rather, the modelling groups have, over a period of many years, shared methods and results from the models.
Finally, the models have never been tested in any out-of-sample, or holdout, fashion. There is therefore no way to evaluate the likely errors caused by the various physical approximations in the models.

None of these reasons would particularly lead one to suspect a warm or cold bias in the models, and so do not give any specific reason to doubt the fundamental conclusion that temperatures will rise in the future.
However, they do make it very difficult to interpret the \emph{spread} of the models, and to understand the likely accuracy of the point forecast that is generated by the mean of the ensemble.
It certainly would not seem appropriate to consider the model predictions as independent random samples from the distribution of possible future climates: because of the lack of independence of the models, and the approximations used in the models, the distribution of future climates is presumably somewhat wider than the spread of the models.
Indeed, the IPCC report itself shows projections with uncertainty wider than the spread of the models
(see figure 10.29).
One way to derive such uncertainties, which we investigate here, would be to assume that the models are \emph{correlated} random samples from the distribution of possible future climates.
We ask the question: given simple assumptions about the correlation structure between the output from the models, what does that tell us about the interpretation of the spread across the models, and how we should use it to derive an
estimate of the real uncertainty?

\section{The Maximum Likelihood Estimator of Variance}

In order to try and answer this question, we now investigate the statistical interpretation of correlated samples. We take a classical, as opposed to Bayesian, statistical approach. This is not for any particular philosophical reason, but just because the principles we want to express (the impact of correlation on estimates of variance) can be captured very simply and elegantly using classical statistics. Bayesian statistics is particularly useful when parameter uncertainty is the focus of attention, but that is not the case here, although it
might nevertheless be useful to repeat this analysis in a Bayesian framework.

Our starting point is that there is an unknown distribution for temperatures in 2050.
Our basic assumptions are that this distribution is normal, and that the IPCC climate model predictions
represent $n$ correlated samples from this distribution, which we write as $x_1, x_2, ..., x_n$.
What information can we derive from these samples about the underlying distribution?
We use the standard classical statistical method of maximum likelihood, due to~\citet{fisher1912}, in which the parameters of the distribution are varied to maximise the probability density of the observed sample,
in order to determine best estimates.

The starting point for our analysis is the probability
density for the multivariate normal distribution, which is given by:

\begin{equation}
p(x|\mu,\Sigma)=
\frac{1}{{(2\pi)}^{n/2}(\mbox{det}\Sigma)^{1/2}}
\mbox{exp} \left( -\frac{1}{2} (x-\mu)^T \Sigma^{-1} (x-\mu) \right)
\end{equation}

where

\begin{itemize}

  \item $x=(x_1,...,x_n)$ is a vector containing the $n$ correlated random samples from the unknown distribution, which are the individual forecasts for temperature in 2050 from the IPCC climate models. $x$ is a \emph{single
      realisation} of the random variable for the multivariate normal we are considering, while, at the same time, the components of $x$ are different predictions for temperature in 2050

  \item $\mu$ is a vector, containing the (unknown) means of this distribution. We assume that all models are samples from the same distribution, so all models have the same mean, and we write $\mu=(\mu,...,\mu)$ (in other words
      we use the symbol $\mu$ for both the vector and the components of the vector).

  \item $\Sigma$ is the (unknown) correlation matrix between the output from the different models.

  \item $\mbox{det} \Sigma$ is the determinant of this correlation matrix

\end{itemize}

We are interested in varying the parameters to maximise $p(x|\mu,\Sigma)$.
This can be done by maximising either $p(x)$ or $\ln p(x)$, and the latter is more convenient, so the next
step is to take the log, which gives:

\begin{equation}
\ln p(x)=-\frac{n}{2}\ln 2\pi -\frac{1}{2}\ln \mbox{det} \Sigma -\frac{1}{2} (x-\mu)^T \Sigma^{-1} (x-\mu)
\end{equation}

The maximum likelihood estimate for the scalar
$\mu$ is $\hat{\mu}=\frac{1}{n} \sum_{i=1}^{n} x_i$. Defining the vectors
$\hat{\mu}=(\hat{\mu},...,\hat{\mu})$ and
$y=x-\hat{\mu}$ then gives:

\begin{equation}
\ln p(x)=-\frac{n}{2}\ln 2\pi -\frac{1}{2}\ln \mbox{det} \Sigma -\frac{1}{2} y^T \Sigma^{-1} y
\end{equation}

Typically at this point one would vary $\Sigma$ to maximise $\ln p(x)$, and hence derive the
maximum likelihood estimator for $\Sigma$. However in our case this is not possible,
since we do not have enough data to estimate $\Sigma$: in fact we only have a single realisation of the random variable
$x$, which gives us just $n$ data points, and $\Sigma$, a symmetric $n$ by $n$ matrix, has many more than
$n$ degrees of freedom. How one might actually go about estimating $\Sigma$ from observed data and new types of model
experiments is discussed in appendix~\ref{a1}.
In this study, however, we will assume a certain structure for $\Sigma$, to reduce the degrees of freedom.
But first, we rewrite the problem slightly by writing the (unknown) variance of the distribution
of temperatures in 2050 as $\sigma^2$, and the (unknown) correlation matrix as $C$, which gives:
\begin{eqnarray}
\Sigma          &=&\sigma^2 C\\
\Sigma^{-1}     &=&\frac{1}{\sigma^2}C^{-1}\\
\mbox{det}\Sigma&=&\sigma^{2n} \mbox{det} C
\end{eqnarray}

$\ln p(x)$ then becomes:

\begin{eqnarray}
\ln p(x)&=&-\frac{n}{2}\ln 2\pi -\frac{1}{2}\ln (\sigma^{2n} \mbox{det} C) -\frac{1}{2\sigma^2} y^T C^{-1} y\\
        &=&-\frac{n}{2}\ln 2\pi -n \ln \sigma-\frac{1}{2}\ln \mbox{det} C -\frac{1}{2\sigma^2} y^T C^{-1} y
\end{eqnarray}

We now make an approximation to the structure of $C$, which is to assume that all the model outputs are correlated
equally with correlation $r$.
This is clearly not likely to be exactly correct, since some pairs of models are presumably more similar
than others, but it serves as a very useful simplification of the problem.
Given this assumption, the correlation matrix for $n=1$ can now be written as:

\begin{equation}
\left(
\begin{array}{c}
  1 \\
\end{array}
\right)
\end{equation}

for $n=2$ as

\begin{equation}
\left(
\begin{array}{cc}
  1 & r \\
  r & 1
\end{array}
\right)
\end{equation}

for $n=3$ as:
\begin{equation}
\left(
\begin{array}{ccc}
  1 & r & r \\
  r & 1 & r \\
  r & r & 1
\end{array}
\right)
\end{equation}

and so on, with $1$'s on the diagonal and $r$'s off the diagonal.

Each of these matrices can also be written as a sum of a multiple of the identity matrix and a multiple of a matrix of ones.
For the $n=3$ case:
\begin{eqnarray}
\left(
\begin{array}{ccc}
  1 & r & r \\
  r & 1 & r \\
  r & r & 1
\end{array}
\right)
&=&
(1-r)\left(
\begin{array}{ccc}
  1 & 0 & 0 \\
  0 & 1 & 0 \\
  0 & 0 & 1
\end{array}
\right)
+r\left(
\begin{array}{ccc}
  1 & 1 & 1 \\
  1 & 1 & 1 \\
  1 & 1 & 1
\end{array}
\right)\\
&=&(1-r) I+r 1
\end{eqnarray}

The inverse of the matrix $C$ is then given by the matrix $\alpha I + \beta 1$
where
\begin{eqnarray}
  \alpha &=& \frac{1}{1-r}\\
  \beta  &=& \frac{r}{(r-1)(1-r+rn)}
\end{eqnarray}

(this is shown in appendix~\ref{a2}).

This gives:
\begin{eqnarray}
y^T C^{-1} y
&=& y^T ( \alpha I + \beta 1) y\\
&=& \alpha y^T I y + \beta y^T 1 y\\
&=& \alpha y^T y + \beta y^T 1 y\\
&=& \alpha y^T y + \beta \left(\sum y_i\right)^2\\
&=&\alpha y^T y
\end{eqnarray}

since $\sum y_i=0$ by definition.

and so $\ln p(x)$ becomes:
\begin{eqnarray}
\ln p(x)
&=&-\frac{n}{2}\ln 2\pi -n \ln \sigma-\frac{1}{2}\ln \mbox{det} C -\frac{\alpha}{2\sigma^2} y^T y
\end{eqnarray}

If we now differentiate with respect to $\sigma$, for fixed $r$, we get:
\begin{eqnarray}
\frac{\partial \ln p(x)}{\partial \sigma}
&=&-\frac{n}{\sigma}+\frac{\alpha}{\sigma^3} y^T y
\end{eqnarray}

solving for $\sigma$ gives:
\begin{eqnarray}\label{eq24}
\hat{\sigma}^2&=&\frac{y^T y}{n(1-r)}
\end{eqnarray}

This is now an estimate for the variance of the unknown distribution of temperature in 2050,
given a certain value for the correlation
between the models. There are two obvious limiting cases. When $r=0$ and the models are independent, we have:
\begin{eqnarray}
\hat{\sigma}^2&=&\frac{y^T y}{n}
\end{eqnarray}
which is the standard maximum likelihood estimator for the variance for independent samples.
When $r=1$ and the models
are fully correlated, the estimated variance is infinite.
In between, for $r=0.5$, we have:
\begin{eqnarray}
\hat{\sigma}^2&=&\frac{2y^T y}{n}
\end{eqnarray}
and we see that  the estimated variance is inflated relative to the independent case, as expected.

\section{Application to UK temperature data}

We now apply equation~\ref{eq24} for the variance of a climate prediction to CMIP3 data for
UK temperature. The data we use is exactly the data used and described in~\citet{hawkins09a}. We use
just one of the emissions scenarios (the A1B scenario), and consider four
of our own different correlation scenarios, with the correlation coefficient $r$ set to 0, 0.25, 0.5 and 0.75.
In fact we have we no real idea what this correlation should be, and maybe it should be as high as, say, 0.99.
A method that could be used to estimate this correlation is discussed in appendix~\ref{a1}.

In figure~\ref{fig1} all four panels show the raw data in grey,
which is the predicted changes in UK temperature for the A1B scenario
for each of the 15 models. The black lines show the ensemble mean of the models.
The red lines then show the predicted uncertainty, calculated using equation~\ref{eq24}, for the four
correlation scenarios.
As expected, increasing the assumed correlation between the models
gradually increases the estimate of the variance of the prediction.

In figure~\ref{fig2}, we only apply the variance adjustment given by equation~\ref{eq24} to the
model to model variability in the ensemble, and not to the internal variability, since it would
be reasonable to assume that the internal variability is uncorrelated between the models. This results
in a lower spread for the same correlation assumption, and it also results in the variance
inflation varying with time (although this isn't obvious from the figure) since the relative sizes of the
model and internal variability components of the total variability change from decade to decade
(see~\citet{hawkins09a}, figure 3).

\section{Discussion}

We have considered the question of how to derive estimates of uncertainty around predictions of future
climate made using multi-model ensembles. There are a number of difficulties with respect to interpreting
the nature of such ensembles, and it is thus difficult to made good predictions of the uncertainty.
We adopt a simple approach whereby we assume that the predictions from the different models are \emph{correlated}
samples from the distribution of future temperatures. We then derive the maximum likelihood estimator of the
unknown variance of the prediction, taking into account the correlation. We find, as expected, that the higher
the assumed correlation, the wider the predicted distribution, according to a very simple formula.

We do not attempt to estimate the correlation, and actually we would consider it effectively impossible to estimate
the correlation at this point. Instead we test out various `correlation scenarios', and consider their impact.
Essentially we have simply replaced the idea of arbitrarily inflating the spread with the idea of arbitrarily
choosing a correlation between the models. Although this clearly does not solve the problem of how to make
accurate probabilistic climate predictions, we do think it is an interesting alternative way to view it.
The only way to actually estimate correlation values from data would be to compare predictions from the various models
in an out-of-sample sense, and see how similar they are to both reality and each other (we discuss this in more
detail in appendix~\ref{a1}).

There are a number of ways one could extend the current piece of work.
Of particular interest to us is to develop methods that can use the uncertainty estimates that we have produced.
Climate forecasts with the same mean, but different levels of uncertainty, should certainly not be treated in the
same way, and understanding that will be crucial to making effective use of climate forecasts in the future.

\appendix

\section{Estimating the model to model correlation matrix}\label{a1}

How \emph{could} the correlation matrix between output from different models be estimated in practice?
One method would be the following:
\begin{itemize}

  \item The $n$ climate models are set up so that their free parameters can be tuned objectively against observational data, using maximum likelihood.

  \item The climate models are tuned to data up to and including 1980, 1985, 1990, 1995, 2000 and 2005, leading to different
  parameter settings in each case.

  \item Each model is then run in forecast mode up to 2010, for a large initial condition ensemble.
  For the model tuned using data up to 1980 that would be
  a 30 year lead-time forecast. For the model tuned using data up to 2005 that would only be a 5 year forecast.
  For 5 year lead-time forecasts this would give 6 examples, for 10 year lead-time forecasts 5 examples, and so on.

  \item The forecasts thus produced can be compared against observational data. Consider the 15 year lead-time forecast, for which there are 4 realisations. One could treat these 4 realisations as four independent
      tests of the ability of each model to predict real climate. For each model, one could derive a bias-correction, and between the models one could derive correlations. These bias corrections and correlations would be very poorly
      estimated, and one would probably want to use a Bayesian approach to account for estimation uncertainty.

  \item Finally, the estimated bias corrections and correlations (whether point estimates or posterior
  distributions) could be applied to derive probabilistic forecasts from real predictions of the period post 2010.

  \item It would also be possible to assess whether these corrections are approximately independent of lead-time: if so then
  it may be possible to apply them to lead times longer than those over which they were developed.
\end{itemize}

This method clearly has some limitations: the paucity of data with which to tune the model prior to 1980, the short
lead times of the predictions, the internal decadal variability in the observed data, and others.
However it has the great advantage of being a genuine out-of-sample evaluation
of the models' predictive performance, which would yield essential information about how to calibrate forecasts of
the future.

\section{Inverse of the correlation matrix}\label{a2}

We now demonstrate that the inverse of the matrix $C=(1-r) I+r 1$ is given by $C^{-1}=\alpha I + \beta 1$
where
\begin{eqnarray}
  \alpha &=& \frac{1}{1-r}\\
  \beta  &=& \frac{r}{(r-1)(1-r+rn)}
\end{eqnarray}

Let $D=\alpha I + \beta 1$.
Then
\begin{eqnarray}
  C D &=& [(1-r) I + r 1] [\alpha I + \beta 1] \\
      &=&  (1-r) \alpha I I + (1-r) \beta I 1 + r \alpha 1 I + r \beta 1 1 \\
      &=&  (1-r) \alpha I + (1-r) \beta 1 + r \alpha 1  + r \beta n 1 \\
      &=&  (1-r) \alpha I + [(1-r) \beta + r \alpha + r n \beta] 1
\end{eqnarray}

Choosing $\alpha=1/(1-r)$ then gives:
\begin{eqnarray}
  C D &=&  I + [(1-r) \beta + r \alpha + r n \beta] 1
\end{eqnarray}

For $D$ to be the inverse of $C$, we must therefore have:
\begin{equation}
(1-r) \beta + r \alpha + r n \beta=0
\end{equation}
which is solved by:
\begin{equation}
  \beta  = \frac{r}{(r-1)(1-r+rn)}
\end{equation}

\bibliography{arxiv}

\begin{figure}[!ht]\begin{center}
\scalebox{0.8}{\includegraphics{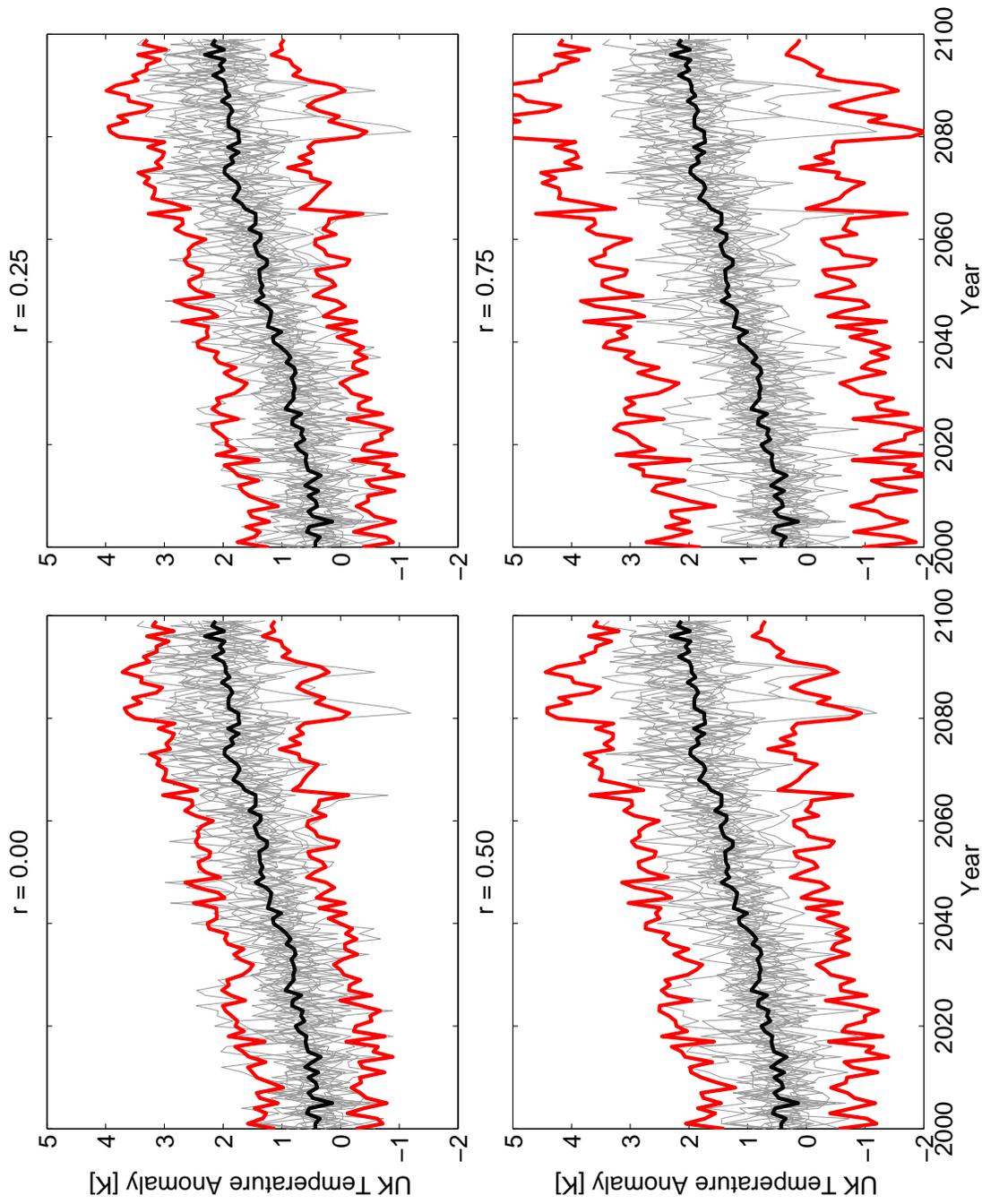}}
\end{center}\caption{
In all panels the grey lines show simulated changes in UK annual mean temperature from CMIP3 data, for the A1B scenario,
for 15 different climate models, and the black line shows the ensemble mean of the models.
The red lines show the estimated uncertainty around the black line (plus and minus two standard deviations),
for four different correlation scenarios for the
correlations between the output from the different models. The correlations used are (a) 0, (b) 0.25, (c) 0.5 and (d) 0.75.
}
\label{fig1}\end{figure}

\begin{figure}[!ht]\begin{center}
\scalebox{0.8}{\includegraphics{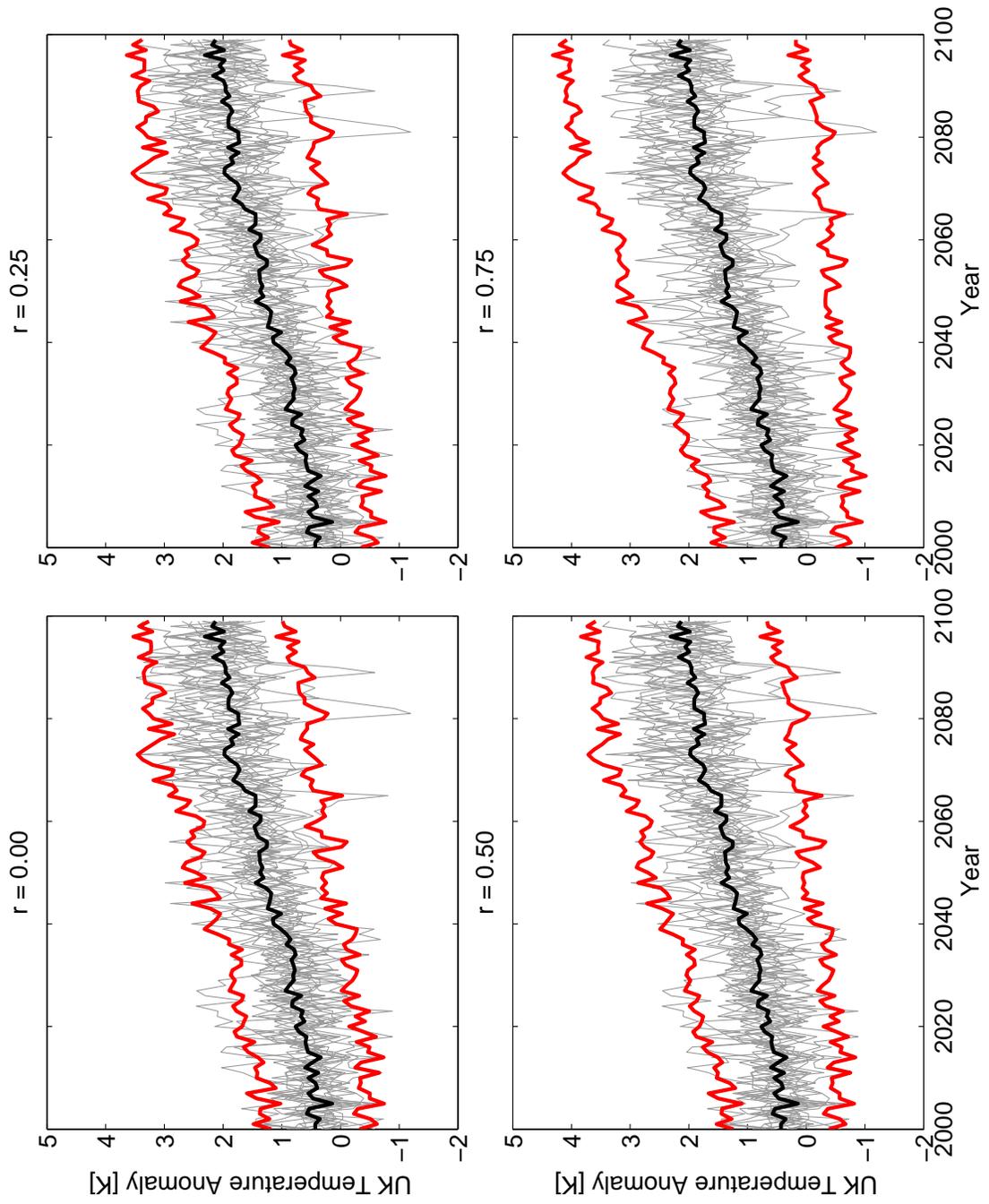}}
\end{center}\caption{
As figure~\ref{fig1}, except that the inflation in the spread due to assumed correlation between the models
is now only applied to the model-to-model variability, not to the internal variability. For fixed correlation
this then results in a lower inflation of the spread.
}
\label{fig2}\end{figure}

\end{document}